\documentclass[11pt,a4paper]{article}

\usepackage{jheppub}
\usepackage{graphicx}
\usepackage{dcolumn}
\usepackage{bm}
\usepackage{amsmath}
\usepackage{braket}
\usepackage{slashed}    
\usepackage{epstopdf}
\usepackage{placeins}
\usepackage{multirow}
\usepackage{makecell}
\usepackage[shortlabels]{enumitem}
\usepackage{cleveref}
\usepackage{xcolor}

\renewcommand\thesection{\arabic{section}}
\newcommand{\beq}{\begin{eqnarray}}
\newcommand{\eeq}{\end{eqnarray}}
\newcommand{\beqnn}{\begin{eqnarray*}}
\newcommand{\eeqnn}{\end{eqnarray*}}
\newcommand{\Tr}{\ensuremath{\mathrm{Tr}}}
\newcommand{\round}{\ensuremath{\mathrm{round}}}
\newcommand{\SU}{\mathrm{SU}}
\newcommand{\YM}{\mathrm{YM}}
\newcommand{\clov}{\mathrm{clov}}

\newcommand{\cool}{\mathrm{cool}}
\newcommand{\tauint}{\tau_{\mathrm{int}}}

\begin{document}

\title{Full QCD with milder topological freezing}

\author[a]{Claudio Bonanno,}
\emailAdd{claudio.bonanno@csic.es}

\author[b]{Giuseppe Clemente,}
\emailAdd{giuseppe.clemente@unipi.it}

\author[b]{Massimo D'Elia,}
\emailAdd{massimo.delia@unipi.it}

\author[c]{Lorenzo Maio,}
\emailAdd{lorenzo.maio@cpt.univ-mrs.fr}

\author[b]{Luca Parente}
\emailAdd{luca.parente@phd.unipi.it}

\affiliation[a]{Instituto de F\'isica Te\'orica UAM-CSIC, c/ Nicol\'as Cabrera 13-15, Universidad Aut\'onoma de Madrid, Cantoblanco, E-28049 Madrid, Spain}

\affiliation[b]{Dipartimento di Fisica dell'Universit\`a di Pisa and INFN - Sezione di Pisa, Largo Bruno Pontecorvo 3, I-56127 Pisa, Italy}

\affiliation[c]{Aix Marseille Univ, Universit\'{e} de Toulon, CNRS, CPT, Marseille, France}

\abstract{We simulate $N_f=2+1$ QCD at the physical point combining open and periodic boundary conditions in a parallel tempering framework, following the original proposal by M.~Hasenbusch for $2d$ $\mathrm{CP}^{N-1}$ models, which has been recently implemented and widely employed in $4d$ $\mathrm{SU}(N)$ pure Yang--Mills theories too. We show that using this algorithm it is possible to achieve a sizable reduction of the auto-correlation time of the topological charge in dynamical fermions simulations both at zero and finite temperature, allowing to avoid topology freezing down to lattice spacings as fine as $a \sim 0.02$ fm. Therefore, this implementation of the Parallel Tempering on Boundary Conditions algorithm has the potential to substantially push forward the investigation of the QCD vacuum properties by means of lattice simulations.}


\maketitle
\flushbottom

\section{Introduction}

The temperature-dependence of the QCD topological susceptibility $\chi = \braket{Q^2}/V$ above the chiral crossover represents an essential theoretical input for axion cosmology. This hypothetical particle was first introduced to solve the strong-CP problem~\cite{Peccei:1977hh,
Peccei:1977ur, Wilczek:1977pj, Weinberg:1977ma}, and then proposed as a possible source of Dark Matter. It couples to QCD topology fluctuations via the axial anomaly, leading in particular to the well-known expression for its effective mass: $m_a^2(T) = \chi(T) / f_a^2$, with $f_a$ the axion decay constant. Under suitable cosmological assumptions, the behavior of $\chi(T)$ up to the GeV scale can be used to put a theoretical bound on $f_a$~\cite{Preskill:1982cy, Abbott:1982af,
Dine:1982ah,Wantz:2009it,Berkowitz:2015aua} (see also Ref.~\cite{DiLuzio:2020wdo}). 

Given that perturbation theory and semi-classical arguments can only provide information about the asymptotically-high temperature limit of $\chi(T)$ via the so-called Dilute Instanton Gas Approximation (DIGA)~\cite{Gross:1980br,Schafer:1996wv,Boccaletti:2020mxu}, lattice QCD simulations represent a natural non-perturbative first-principle tool to study the temperature-dependence of the susceptibility in the chiral-symmetric phase, and this quantity has been the target of several lattice studies in recent years~\cite{Bonati:2015vqz, Petreczky:2016vrs,
Frison:2016vuc, Borsanyi:2016ksw, Bonati:2018blm, Burger:2018fvb,Lombardo:2020bvn, Chen:2022fid,
Athenodorou:2022aay}. Despite most of the available lattice calculations, all exploiting quite different numerical strategies and/or fermion discretizations, find that $\chi(T)$ is well-described by a power-law suppression above \mbox{$T\gtrsim 300\text{ MeV} \sim 2 T_c$}, with \mbox{$T_c \simeq 155$ MeV} the QCD chiral crossover temperature, with an exponent which is qualitatively in agreement with the one predicted by the DIGA, a general consensus among the actual values of $\chi(T)$, especially for $T_c \lesssim T \lesssim 300$ MeV, has not been reached yet (see also Sec.~6 of Ref.~\cite{Aarts:2023vsf} for a recent review on this topic). This issue thus demands further investigations.

From the computational point of view, the determination of $\chi$ on the lattice in the presence of dynamical fermions at finite-temperature is a highly non-trivial numerical challenge, as several computational problems have to be addressed. In particular, a serious limitation is introduced by the infamous \textit{topological freezing} problem, which typically affects the local algorithms customarily employed in lattice field theory simulations. For example, when employing the customary Hybrid Monte Carlo (HMC) algorithm to update the QCD gauge configuration during the Monte Carlo, this algorithm experiences a severe growth in the auto-correlation time $\tau$ of the topological charge $Q$, i.e., the number of updating steps necessary to generate two decorrelated extractions of $Q$, when approaching the continuum limit. More precisely, the growth of $\tau$ is compatible with an exponential behavior in the inverse lattice spacing~\cite{Alles:1996vn,DelDebbio:2004xh,Schaefer:2010hu}. In practice, this means that, if the lattice spacing is taken to be too fine, the Monte Carlo Markov chain tends to remain trapped in a fixed topological sector, and very few to no fluctuations of $Q$ are observed during an affordable simulation, hence the name ``topological freezing''. This loss of ergodicity ultimately prevents from correctly sampling the topological charge distribution unless an unfeasibly large statistics is collected, making it very hard to reliably compute topological quantities on fine lattices. Moreover, it can also possibly introduce undesired biases in other non-topological observables computed on the lattice.

It is important to stress that topological freezing is not a specific problem of the HMC algorithm, hence of full QCD simulations, as it generally affects the numerical simulations performed with customary local algorithms of several other lattice field theories whose continuum counterpart is characterized by the classification of path integral configurations into homotopy classes, spanning from simple one-dimensional models to $4d$ $\SU(N)$ pure Yang--Mills theories, freezing being exponential with $N$ in this case.

For what concerns the lattice calculation of the topological susceptibility in full QCD, being able to simulate very fine lattice spacings is an essential necessity for two different reasons. On one hand, it is well known that, when adopting a non-chiral discretization of the sea quarks (such as Wilson or staggered), the topological susceptibility suffers for large lattice artifacts which can exactly be traced back to the explicit breaking of the chiral symmetry of the adopted lattice-discretized Dirac operator, in particular to the absence of exact zero modes which results in the lack of a proper suppression of topological fluctuations. At the practical level, performing reliable continuum extrapolations using lattice spacings $a \gtrsim 0.06$~fm is typically very hard,
and large systematic errors will affect the final results. On the other hand, finite-temperature calculations are performed on lattices where the temporal extent $N_t$ is smaller than the spatial one $N_s$, and the temperature is given by the inverse of the temporal size of the lattice $T=1/(a N_t)$, with $a$ the lattice spacing. In this case, in order to obtain
reliable continuum extrapolations, one has to consider
$N_t=10-16$, then it is clear that, to reach the GeV scale, very fine lattice spacings of the order of $a \sim 0.01$ fm are needed. Since simulations for such fine lattice spacings are heavily affected by topological freezing, it is practically very difficult to reach temperatures above $\sim 500-600$ MeV.

A general exact solution to solve topological freezing is not known, except for ones specifically tailored for toy models~\cite{Bonati:2017woi,Iannelli_thesis,Albandea:2024fui}, but in recent years several different strategies have been devised in the literature to mitigate topological freezing, and to reduce its severity from an exponential down to a polynomial growth with a small power~\cite{Bietenholz:2015rsa,Laio:2015era,Luscher:2017cjh,Giusti:2018cmp,Florio:2019nte,Funcke:2019zna,Kanwar:2020xzo,Nicoli:2020njz,Albandea:2021lvl,Cossu:2021bgn,Borsanyi:2021gqg,papamakarios2021,Fritzsch:2021klm,Abbott:2023thq,Eichhorn:2023uge,Howarth:2023bwk,Bonanno:2024udh} (see also Refs.~\cite{Finkenrath:2023sjg,Boyle:2024nlh} for recent reviews). However, most of these available algorithmic improvements have been applied either to lower-dimensional toy models or to $4d$ pure-gauge theories, with the notable exception of Open Boundary Conditions (OBC) simulations~\cite{Luscher:2011kk,Luscher:2012av}, which have been employed extensively in zero-temperature QCD simulations with dynamical fermions.

The present investigation deals exactly with the problem of mitigating topological freezing in QCD simulations with dynamical quarks on fine lattices. Our main goal is to present a novel implementation of the algorithmic setup put forward in Ref.~\cite{Hasenbusch:2017unr} by M.~Hasenbusch for $2d$ $\mathrm{CP}^{N-1}$ models, and recently implemented also for $4d$ $\mathrm{SU}(N)$ gauge theories in Ref.~\cite{Bonanno:2020hht}, namely the \emph{Parallel Tempering on Boundary Conditions} (PTBC) algorithm. Since in the last few years this algorithm was extensively applied to improve the state of the art of several topological and non-topological observables both in $2d$ $\mathrm{CP}^{N-1}$ models~\cite{Berni:2019bch,Bonanno:2022hmz} and in $4d$ $\mathrm{SU}(N)$ pure-gauge theories at zero~\cite{Bonanno:2020hht,Bonanno:2022yjr,DasilvaGolan:2023cjw,Bonanno:2024ggk,Bonanno:2024nba} and non-zero temperature~\cite{Bonanno:2023hhp} (see also~\cite{Abbott:2024mix} for a recent application of the PTBC algorithm with normalizing flows in the $4d$ $\SU(3)$ pure-gauge theory), it is a natural development to investigate its efficiency in the more computationally demanding case of full QCD simulations.

The underlying idea of this algorithm consists of simulating several replicas of the lattice at the same time, each one differing from the others for the boundary conditions imposed on a few links, chosen so as to interpolate among Periodic Boundary Conditions (PBC) and OBC. Each replica is updated independently adopting standard algorithms (such as the HMC) and, after each update, swaps of configurations among different replicas are proposed via a Metropolis step. Thanks to the swaps, a given gauge configuration experiences different boundary conditions, and the net effect is to ``transfer'' the fast decorrelation of the topological charge achieved thanks to OBC towards the PBC one, where the measure of all the desired observables is always performed.

The PTBC algorithm, thus, allows to have the best of both worlds. On the one hand, the auto-correlation time of $Q$ in the periodic replica is largely reduced by exploiting the advantages of OBC. On the other hand, keeping periodic boundaries for the calculation of physical observables allows us to avoid the unphysical effects introduced by the open boundaries, allowing for a better control of systematic finite-volume effects. As an example, when OBC are considered, since a notion of global topological charge is inevitably lost, one is forced to integrate the topological charge density 2-point correlator to calculate $\chi$. Such integration, however, has to be carefully conducted up to a maximum distance which must lay well within the bulk, so as to safely stay sufficiently far from the boundaries. This thus typically requires much larger lattices to keep finite-size effects under control.

This manuscript is organized as follows. In Sec.~\ref{sec:setup}. we present our algorithmic setup in the zero and in the finite-temperature cases. In Sec.~\ref{sec:res} we present results obtained with the PTBC algorithm, and we compare them with those obtained using the standard one. Finally, in Sec.~\ref{sec:conclu} we draw our conclusions and discuss possible future outlooks of our study.

\section{Lattice setup and implementation}\label{sec:setup}

This section is devoted to discuss our lattice setup, with particular emphasis on our novel algorithmic proposal. It is worth stressing that in the following we will adopt staggered fermions, but this choice does not introduce any loss of generality, and our algorithmic setup can be applied \emph{verbatim} to other fermionic discretizations such as, e.g., the Wilson one.

\subsection{Lattice QCD action}\label{sec:lat_disc}

This section is devoted to discuss how we discretized $N_f = 2+1$ QCD on a $N_s^3 \times N_t$ lattice with lattice spacing $a$ and standard periodic boundary conditions for the gauge fields in all directions. The other replicas with different boundary conditions will be discussed in the next section.

We adopt the tree-level Symanzik-improved gauge action for the gluonic sector, and $2+1$ flavor of rooted stout-smeared staggered fermions for the quark sector. The lattice QCD partition function takes the general form using the discretized path-integral formalism:
\beq
Z_{\rm LQCD} = \int [dU] e^{-S_{\YM}[U]} \det\left\{\mathcal{M}^{(\rm stag)}_{l}[U]\right\}^{\frac{1}{2}} 
\det\left\{\mathcal{M}^{(\rm stag)}_{s}[U]\right\}^{\frac{1}{4}},
\eeq
with $[dU]$ the gauge-invariant Haar measure.

The staggered Dirac operator $D_{\rm stag}$ is
expressed in terms of the stout-smeared gauge links $U^{(2)}$~\cite{Morningstar:2003gk}, obtained after the application of $n_{\rm stout} = 2$ levels of isotropic smearing with stouting parameter $\rho_{\rm stout} = 0.15$:
\begin{gather}
\mathcal{M}^{(\rm stag)}_f[U] \equiv D_{\rm stag}[U^{(2)}] + am_f,\nonumber\\
\label{eq:stag_operator}
D_{\rm stag}[U^{(2)}] = \sum_{\mu=1}^{4}\eta_{\mu}(x)\left( U^{(2)}_{\mu}(x) \delta_{x,y-\hat{\mu}} - {U_{\mu}^{(2)}}^{\dagger}(x-\hat{\mu}) \delta_{x,y+\hat{\mu}} \right)\ ,\\ 
\eta_{\mu}(x) = (-1)^{x_1+\dots+x_{\mu-1}}\ , \nonumber
\end{gather}
where $f=l,s$ stands for, respectively, the light and strange quark flavors.
The gauge action is instead a function of the non-stouted gauge links and is defined as:
\beq
S_{\YM}[U] = - \frac{\beta}{3} \sum_{x, \mu \ne \nu} \left\{ \frac{5}{6}\Re\Tr\left[\Pi_{\mu\nu}^{(1\times1)}(x)\right] 
- \frac{1}{12}\Re\Tr\left[\Pi_{\mu\nu}^{(1\times2)}(x)\right]\right\}\ ,
\eeq
with $\Pi^{(n\times m)}_{\mu\nu}(x)$ the $n\times m$ Wilson loops in the $(\mu,\nu)$ plane centered in the site $x$, and $\beta=6/g^2$ the inverse bare gauge coupling.

Finally, the topological charge is discretized using the simplest definition which is odd under parity inversions, the \emph{clover} discretization:
\beq\label{eq:clov_charge}
Q_{\clov} = \frac{-1}{2^9 \pi^2}\sum_{x}\sum_{\mu\nu\rho\sigma=\pm1}^{\pm4}\varepsilon_{\mu\nu\rho\sigma}
\Tr\left\{\Pi_{\mu\nu}^{(1\times1)}(x)\Pi_{\rho\sigma}^{(1\times1)}(x)\right\}\ ,
\eeq
where it is understood that
$\varepsilon_{\mu\nu\rho\sigma}=-\varepsilon_{(-\mu)\nu\rho\sigma}$. Since such definition would lead to a lattice definition of the topological susceptibility suffering from additive and multiplicative renormalizations~\cite{Campostrini:1988cy,Vicari:2008jw}, it is customary to compute it after smoothing to obtain a fully renormalized observable.

Several different smoothing algorithms have been proposed, such as
cooling~\cite{Berg:1981nw,Iwasaki:1983bv,Itoh:1984pr,Teper:1985rb,Ilgenfritz:1985dz,Campostrini:1989dh,Alles:2000sc},
smearing~\cite{APE:1987ehd, Morningstar:2003gk} and gradient
flow~\cite{Luscher:2009eq, Luscher:2010iy}, all giving perfectly consistent and agreeing results when matched among each other~\cite{Alles:2000sc, Bonati:2014tqa, Alexandrou:2015yba}. In this work we adopt cooling for its simplicity and numerical cheapness. Since, after cooling, the lattice gluonic charge is close to integer values, we adopt the following integer definition~\cite{DelDebbio:2002xa,Bonati:2015sqt}:
\beq
Q = \round\left\{\alpha \, Q_{\clov}^{(\cool)}\right\}\ , \qquad a^4\chi=\frac{\braket{Q^2}}{N_s^3 N_t},
\eeq
with $\alpha=\min_{1 < x < 2} \left\langle\left[x \, Q_{\clov}^{(\cool)} - \round\left\{x \, Q_{\clov}^{(\cool)}\right\}\right]^2\right\rangle$ minimizing the distance between the peaks of the distribution of the cooled charge and integer values, and with $\round(x)$ denoting the closest integer to $x$. Typical values of $\alpha$ turn out to be $\sim 1.03-1.05$, and we employed in all cases 100 steps of cooling, as this is a sufficient number to observe a plateau in $\chi$ as a function of the number of cooling steps in all explored cases.

\subsection{The PTBC algorithm with dynamical fermions}

We now consider $N_r$ replicas of the lattice theory described in the previous section, which differ among them only for the boundary conditions imposed on the links crossing orthogonally a cubic $L_d \times L_d \times L_d$ sub-region of the lattice  placed along one of the spatial boundaries, which we call \emph{the defect}.

We choose to alter the boundary conditions only for the gauge links entering the pure-gauge action, i.e., we choose to leave both the links entering the fermion determinant, and the determinant itself, out from the tempering of the boundary conditions. This is a perfectly legitimate choice, given that the non-periodic replicas are unphysical, and thus their action can be chosen at one's own convenience, and it can be physically justified as follows.

At finite lattice spacing, there is, strictly speaking, no proper notion of topological sectors. However, as the continuum limit is approached, the free-energy barriers between different pseudo-topological sectors grow, eventually diverging and properly restoring disconnected topological sectors. This is, in a few words, the physical origin of topological freezing~\cite{DelDebbio:2004xh}. Our expectation is that these free energy barriers are developed essentially by the gauge action, while the Dirac term should develop at most a step when jumping between two pseudo-topological sectors, associated to a non-zero eigenvalue becoming very small (due to the index theorem). For this reason, we do not expect the algorithmic efficiency of the PTBC algorithm to improve if the boundary conditions of the links entering the fermion determinant are altered, hence our choice.\footnote{Our choice is similar in spirit to one taken in the original PTBC pure-Yang--Mills $\mathrm{SU}(N)$ paper~\cite{Bonanno:2020hht} for what concerns the addition of the $\theta$-term in parallel tempering simulations. In that study, only the standard $\theta=0$ Wilson gauge action took part in the tempering of boundary conditions, while the imaginary-$\theta$ term did not.} Moreover, as we will see in a moment, it will simplify the calculation of the swap probability.

At the practical level, as already done in~\cite{Bonanno:2020hht}, the change of the boundary conditions of specific links is easily achieved by attaching to each link a numerical factor of the form:
\beq
K_{\mu}^{(r)}(x) =
\begin{cases}
c(r), \qquad &\mu=1, \qquad x_1=N_s-1, \qquad 0 \le x_0,x_2,x_3 < L_d,\\
\\
1,    \qquad &\text{elsewhere},
\end{cases}
\eeq
with $0\le c(r) \le 1$. This factor is used to depress the gauge interaction of the links crossing the defect when $c<1$. Eventually, when $c=0$, the coupling of the links crossing the defect vanishes and so does their related force, thus realizing OBC. The case $c=1$, instead, realizes PBC, and identifies the physical replica where measures of $Q$, according to the discretization and procedure described in Sec.~\ref{sec:lat_disc}, are performed.

In the end, it is useful to introduce the following partition function, which depends on the replica index $r$:
\beq
Z_{\rm LQCD}^{(r)} = \int [dU_r] e^{-S_{\YM}^{(r)}[U_r]} \det\left\{\mathcal{M}^{(\rm stag)}_{l}[U_r]\right\}^{\frac{1}{2}} 
\det\left\{\mathcal{M}^{(\rm stag)}_{s}[U_r]\right\}^{\frac{1}{4}},
\eeq
where $U_r$ stands for the gauge configuration of the $r^{\rm th}$ replica, and where
\beq
\begin{aligned}
S_{\YM}^{(r)}[U_r] = - \frac{\beta}{3} \sum_{x, \mu \ne \nu} &\left\{ \frac{5}{6}\mathcal{K}_{\mu\nu}^{(1\times 1)}(x; r)\Re\Tr\left[\Pi_{\mu\nu}^{(1\times1)}(x; r)\right]\right.
\\ &\left. \,\,\,\,- \frac{1}{12}\mathcal{K}_{\mu\nu}^{(1\times 2)}(x; r)\Re\Tr\left[\Pi_{\mu\nu}^{(1\times2)}(x; r)\right]\right\}\ .
\end{aligned}
\eeq
The factors $\mathcal{K}_{\mu\nu}^{(1\times 1)}(x; r)$ and $\mathcal{K}_{\mu\nu}^{(1\times 1)}(x; r)$ are just the products of the factors $K_{\mu}^{(r)}(x)$ along the lattice paths specifying the $1 \times 1$ and $1 \times 2$ plaquettes.

Each replica is updated independently and simultaneously using the standard Rational Hybrid Monte Carlo
(RHMC) algorithm~\cite{Clark:2006fx,Clark:2006wp}. After the update of all the replicas, swaps among two adjacent replicas $(r,s=r+1)$ are proposed sequentially (randomly alternating with equal probability the start of the swap proposals from the $r=0$ and the $r=N_r-1$ replicas), and accepted via a standard Metropolis step:
\beq\label{eq:DeltaActions}
\begin{aligned}
p(r,s) &= \min\left\{1, e^{-\Delta S^{(r,s)}_{\rm swap}}\right\},\\
\\
\Delta S^{(r,s)}_{\rm swap} &= S_{\YM}^{(r)}[U_s]+S_{\YM}^{(s)}[U_r]-S_{\YM}^{(r)}[U_r]-S_{\YM}^{(s)}[U_s].
\end{aligned}
\eeq
As earlier anticipated, since the fermion determinant does not participate to the tempering on boundary conditions, the variation of the action only involves the gauge sector, and just requires the calculation of a few terms, since the contribution to $\Delta S^{(r,s)}_{\rm swap}$ of links which are more then 2 lattice spacings far from the defect exactly vanishes. Given that the optimal setup is achieved when the mean swap acceptances $P_r \equiv \braket{p(r,r+1)}$ are roughly constant, we performed short test runs to tune the $c(r)$ parameters to achieve an approximately constant value of $P_r$ among the different replicas: $P_r \approx P \approx $ constant.

Finally, after each update and each swap proposal, the periodic replica, which is translation-invariant, is translated by one lattice spacing in a random direction. This step is done to effectively move the position of the defect, and is expected to improve the efficiency of the algorithm, as moving the defect corresponds to moving the position where topological excitations are more likely to be created/annihilated.

\subsection{Multicanonical PTBC simulations at finite-temperature}\label{sec:multican}

When performing finite-temperature simulations, a further computational issue regarding topological excitations, unrelated to freezing, arises. As explained in the Introduction, the topological susceptibility is rapidly suppressed above the chiral crossover, meaning that, for the typical volumes employed in lattice simulations,
\beq
\braket{Q^2} = \chi V \ll 1.
\eeq
This means that the topological charge distribution is largely dominated by $Q=0$, and topological excitations are extremely rare. We stress that this problem is not a computational issue related to the loss of ergodicity of the RHMC algorithm close to the continuum limit, but rather a physical effect due to the suppression of the susceptibility at high temperatures, which thus appears on top of topological freezing.

In order to overcome this further complication, we employ, on top of the PTBC algorithm, \emph{multicanonical simulations}. This numerical strategy, first introduced to study strong first-order phase transitions~\cite{Berg:1992qua}, has been, in recent times, successfully applied to improve the sampling of rare topological fluctuations in simulations characterized by extremely small values of $\braket{Q^2}$~\cite{Bonati:2017woi, Jahn:2018dke, Bonati:2018blm, Athenodorou:2022aay,Bonanno:2022dru}. In this work we will use the same multicanonical setup of Refs.~\cite{Bonati:2018blm, Athenodorou:2022aay}, which can be easily included in the PTBC setup outlined in the previous section. We briefly summarize it in the following.

The underlying idea is to add to the gauge action a bias potential:
\beq
S_{\YM}^{(r)}[U_r] \longrightarrow S_{\YM}^{(r)}[U_r] + V_{\rm topo}\left(Q_{\rm mc}[U_r]\right),
\eeq
where the shape of $V_{\rm topo}$ is specifically chosen to improve the probability of visiting suppressed topological sectors, without spoiling importance sampling. In particular, one needs to avoid a potential which is too strong, as it could induce an overlap problem; this can be easily avoided, and we rely on the same potential choices of the original references~\cite{Bonati:2018blm, Athenodorou:2022aay}, which are shown to introduce no pathological behavior.

Clearly, expectation values $\braket{\mathcal{O}}$ with respect to the original path-integral distribution are exactly recovered through a standard reweighting procedure:
\beq\label{eq:reweighting_formula}
\braket{\mathcal{O}} = \frac{\braket{\mathcal{O}e^{V_{\rm topo}(Q_{\rm mc})}}_{\rm mc}}{\braket{e^{V_{\rm topo}(Q_{\rm mc})}}_{\rm mc}}\ ,
\eeq
with $\braket{\cdot}_{\rm mc}$ standing for the expectation value computed in the presence of $V_{\rm topo}$.

Concerning the argument of $V_{\rm topo}$, the quantity $Q_{\rm mc}$ stands for some gluonic discretization of the topological charge, which does not need to be the same used for the computation of the susceptibility previously introduced; in particular, while on one side it is desirable that a good correlation
exists between the latter and $Q_{\rm mc}$, so that the potential can effectively modify the 
sampling of topological sectors, on the other side it is advisable that $Q_{\rm mc}$ is not 
too close to integer values, as that could induce a step-like
behaviour in the molecular dynamics evolution of the bias potential, which could lead to 
the necessity of extremely small integration steps in the HMC. In our simulations, we will use the clover discretization~\eqref{eq:clov_charge} computed after a few steps of stout-smearing (in this paper we employed $10$ stout-smearing steps with stout parameter $\rho_{\rm mc}=0.1$ to define $Q_{\rm mc}$). This choice is particularly convenient because it makes $Q_{\rm mc}$ analytically-dependent on the non-smeared links, thus allowing us to easily use the same RHMC algorithm also in the presence of the topological potential.

As a final note, we observe that, in this case, we will use the same topological potential for all the replicas, meaning the multicanonical term does not participate in the tempering of boundary conditions, and thus does not enter in the calculation of the swap probability either.

\subsection{Parallelization on a multi-GPU device}\label{parallelization}

Finally, let us spend a few words on the practical implementation of the numerical code running
on a multi-GPU device, in particular on the \href{https://leonardo-supercomputer.cineca.eu/hpc-system/}{\texttt{Leonardo}} machine at \href{https://www.cineca.it/en/about-us/organization}{CINECA}. 
This is an important aspect since, as an additional benefit of the new algorithm,
which is shared with other parallel tempering algorithms, part of the parallelization reveals trivial
and highly efficient,
with different copies running independently on different portions of the machine. Indeed, our implementation extends a former multi-node and multi-GPU distributed code (publicly available in~\cite{openStapleCode}) where the RHMC algorithm applies to a
single lattice, partitioned in one of the space-time directions (usually the longest one) into multiple MPI ranks associated with different devices, as described in Refs.~\cite{Bonati:2017ovw,Bonati:2018wqj}.
In our parallel implementation of the parallel tempering algorithm, each copy of the lattice runs the same RHMC operations of the code without replicas in a single instruction multiple data (SIMD) fashion, with the only difference lying in the boundary conditions at the defect, which vary depending on the replica label.
This parallelization becomes straightforward thanks to the use of independent MPI communicators and groups for each lattice copy, therefore guaranteeing communications only between devices associated with sub-lattices that share borders during the RHMC steps, mimicking the former implementation without replicas
which was already introduced in~\cite{Bonati:2017ovw,Bonati:2018wqj}.
However, since shared memory is not available when dealing with larger lattices and numbers of replicas, instead of swapping whole configurations between different devices as discussed above, our practical choice is to swap only the boundary conditions at the defect, which involves only a small amount of data. Therefore the labels identifying specific replicas are also swapped accordingly between different MPI ranks. Finally, the whole coordination for the swaps tests is managed by a single MPI master rank, which gathers action differences from each replica pair and computes acceptances according to Eq.~\eqref{eq:DeltaActions}.

\section{Results}\label{sec:res}

This section is devoted to the discussion the results obtained with the PTBC algorithm, both at zero- and finite-temperature, and to the comparison them with the results obtained with the standard RHMC algorithm, which in the following will be just addressed as ``the standard algorithm'' for brevity. We recall that multicanonical simulations were employed only for our finite-temperature runs.

\subsection{Zero-temperature results}

We start our investigation from the zero-temperature case, where the value of the QCD topological susceptibility for physical quark masses can be reliably
computed using Chiral Perturbation Theory (ChPT)~\cite{DiVecchia:1980yfw,Leutwyler:1992yt,Mao:2009sy,Guo:2015oxa,GrillidiCortona:2015jxo,Luciano:2018pbj}, which can thus be used as a useful benchmark to be compared with lattice results. At Next-to-Leading Order (NLO), and for two degenerate light quark flavors $m_u = m_d = \left(m_u^{(\rm phys)}+m_d^{(\rm phys)}\right)/2$, one obtains~\cite{GrillidiCortona:2015jxo,Bonati:2015vqz}:
\beq\label{eq:chi_T0_ChPT}
\chi^{1/4}=77.8(4)\text{ MeV}, \qquad \text{ (NLO ChPT)}.
\eeq

This result has been checked thoroughly by lattice simulations~\cite{Bonati:2015vqz,Borsanyi:2016ksw,Athenodorou:2022aay}. In Ref.~\cite{Bonati:2015vqz}, a lattice spacing range corresponding to $a \sim 0.12 - 0.057$ fm was explored, and significant lattice artifacts were found for the smoothed gluonic discretization. In particular, for the finest lattice spacing explored, the lattice susceptibility was found to be still more than 2 times larger than its expected continuum value. A continuum extrapolation of the lattice data yields a compatible result with the ChPT prediction, although systematics related to the continuum extrapolation were sizable. Compatible extrapolations using similar lattice spacing ranges were found in Refs.~\cite{Borsanyi:2016ksw,Athenodorou:2022aay} using two rather different methods to reduce the size of lattice artifacts affecting the susceptibility. In Ref.~\cite{Borsanyi:2016ksw}, the authors employed an \emph{ad hoc} reweighting method based on the lowest-lying eigenvalues of the staggered Dirac operator. In Ref.~\cite{Athenodorou:2022aay}, instead, a fermionic definition of $Q$ based on the index theorem and relying on the low-lying spectrum of the Dirac operator was adopted~\cite{Giusti:2008vb,Bonanno:2019xhg}. 

We thus performed simulations for two fairly fine lattice spacings $a=0.0480$ fm and $a=0.0397$ fm at zero temperature, both using the standard and the PTBC algorithm. Our goal is first to compare the performances of these two algorithms, and then to compare our results for $\chi$ with the ChPT prediction and with previous lattice determinations. We expect that our determinations for these two fine lattice spacings should exhibit small lattice artifacts, and thus be reasonably close both to the ChPT result as well as to the continuum extrapolations reported in the literature.

For our $T=0$ simulations we considered hypercubic $N_s^4$ lattices and chose the bare lattice parameters so as to follow a Line of Constant Physics (LCP) with physical quark masses, i.e., physical pion mass $m_\pi=135$ MeV and physical strange-to-light quark mass ratio $m_s/m_l=28.15$. The summary of the simulation parameters is reported in Tab.~\ref{tab:simul_pars_zero_T}, along with the details of the PTBC setup. The volume is in both cases larger than $1.5$ fm, and is thus expected to be sufficiently large to be insensitive to finite-size effects within our typical statistical errors~\cite{Bonati:2015vqz}.

\begin{table}[!t]
\begin{center}
\begin{tabular}{|c|c|c|c|c|c|c|}
\hline
$\beta$ & $a$~[fm] & $am_s \cdot 10^{2}$ & $N_s$ & $N_r$ & $L_d$ & $P$ $(\%)$\\
\hline
\multirow{3}{*}{4.2512} & \multirow{3}{*}{0.0480} & \multirow{3}{*}{1.891} & \multirow{3}{*}{48} & 10 & \multirow{2}{*}{2} & 21(4)\\
 &  &  & & 13 & & 32(3)\\
  &  &  & & 24 & 4 & 20(4)\\
 \hline
4.3600 & 0.0397 & 1.590 & 48 & 13 & 2 & 30(4)\\
\hline
\end{tabular}
\end{center}
\caption{Summary of simulation parameters for the $T\simeq 0$ runs, performed on
hypercubic $N_s^4$ lattices. The bare parameters $\beta$ and $am_s$ and
the lattice spacings $a$ have been fixed according to the LCP determined in Refs.~\cite{Aoki:2009sc, Borsanyi:2010cj, Borsanyi:2013bia} with $m_\pi = 135$ MeV, and
$am_l$ is fixed through the physical strange-to-light quark mass ratio $m_s/m_l=28.15$. The value of $P$ represents the average value of the swap probabilities, and the dispersion around it.}
\label{tab:simul_pars_zero_T}
\end{table}

\begin{figure}[!t]
\centering
\includegraphics[scale=0.42]{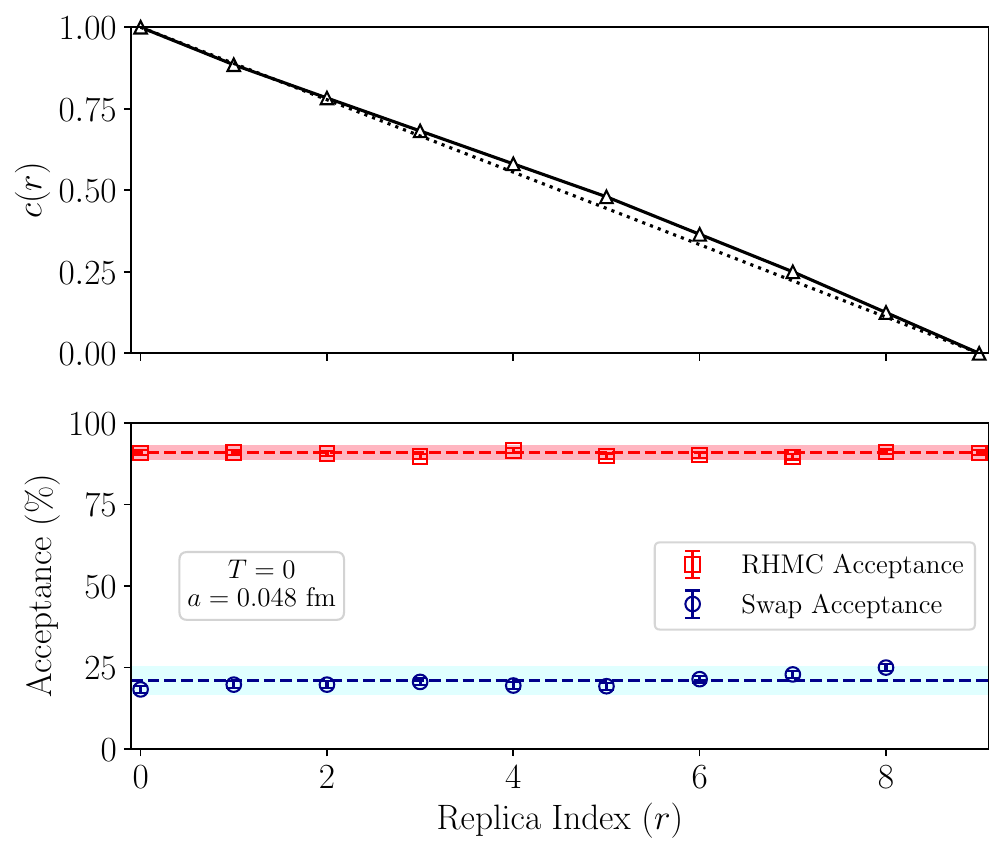}
\includegraphics[scale=0.42]{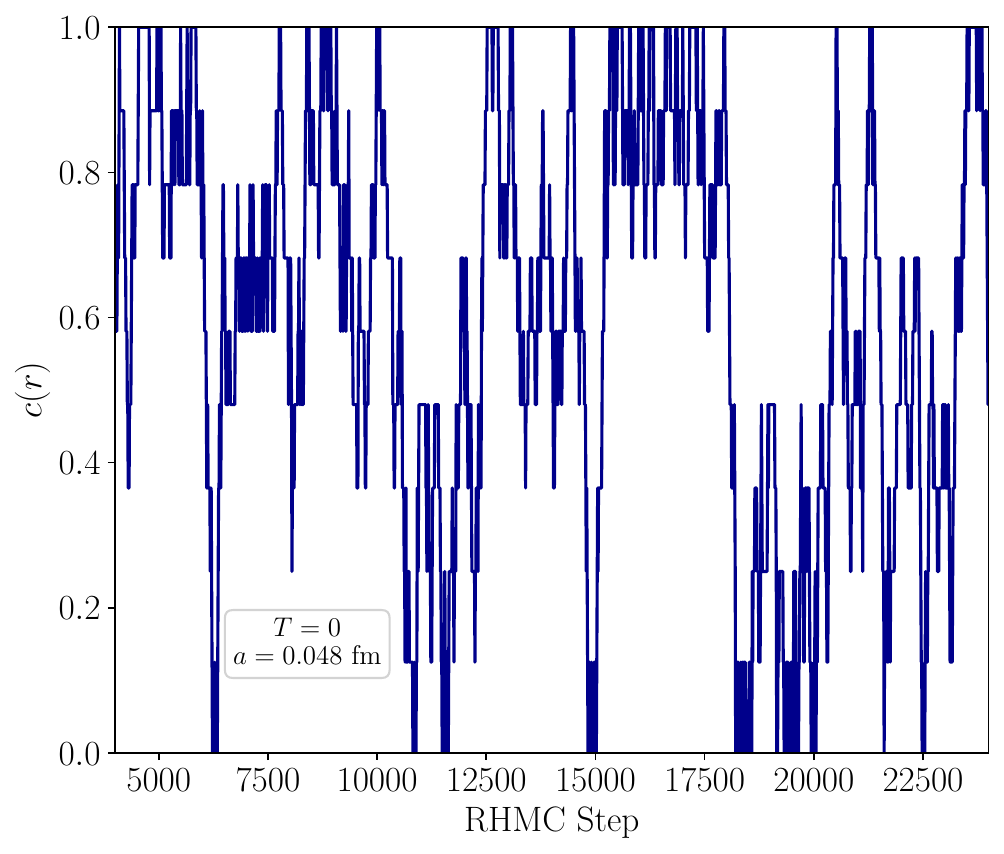}
\caption{These figures refer to the $T=0$ simulation point with $a=0.0480$ fm. Top left: employed values of $c(r)$ compared to a simple linear behavior (dotted line). Bottom left: swap acceptances $P_r=\braket{p(r,r+1)}$ and RHMC acceptances as a function of the replica index $r$. In this case, we find that $P_r \approx P = 21(3) \%$, where the error stands for the dispersion of the acceptances around their average value. Also, the RHMC acceptances are fairly constant among the different replicas, as we find $P_{\rm RHMC} \approx 91(2) \%$. Right: evolution of the boundary conditions experienced by a given configuration followed during its Monte Carlo evolution.}
\label{fig:acc}
\end{figure}

Let us start by discussing some details of the PTBC setup. In Fig.~\ref{fig:acc}, we show an example for the simulation point with $a=0.048$ fm of the calibrated values of the $c(r)$ tempering parameters, tuned through short test runs to achieve an approximately constant swap acceptance rate among adjacent replicas $P_r = \braket{p(r,r+1)}\approx P$. When following a given configuration during its Monte Carlo evolution through the various replicas, this allows it to uniformly explore different values of $c(r)$, ensuring the correct behavior of the parallel tempering. We also observe that keeping the same integration step size across the various replicas yields the same RHMC acceptance rates in all cases, which was always of the order of $\sim 90 \%$, meaning that no tuning is required for what concerns molecular dynamics parameters.

\newpage

\begin{figure}[!t]
\centering
\includegraphics[scale=0.42]{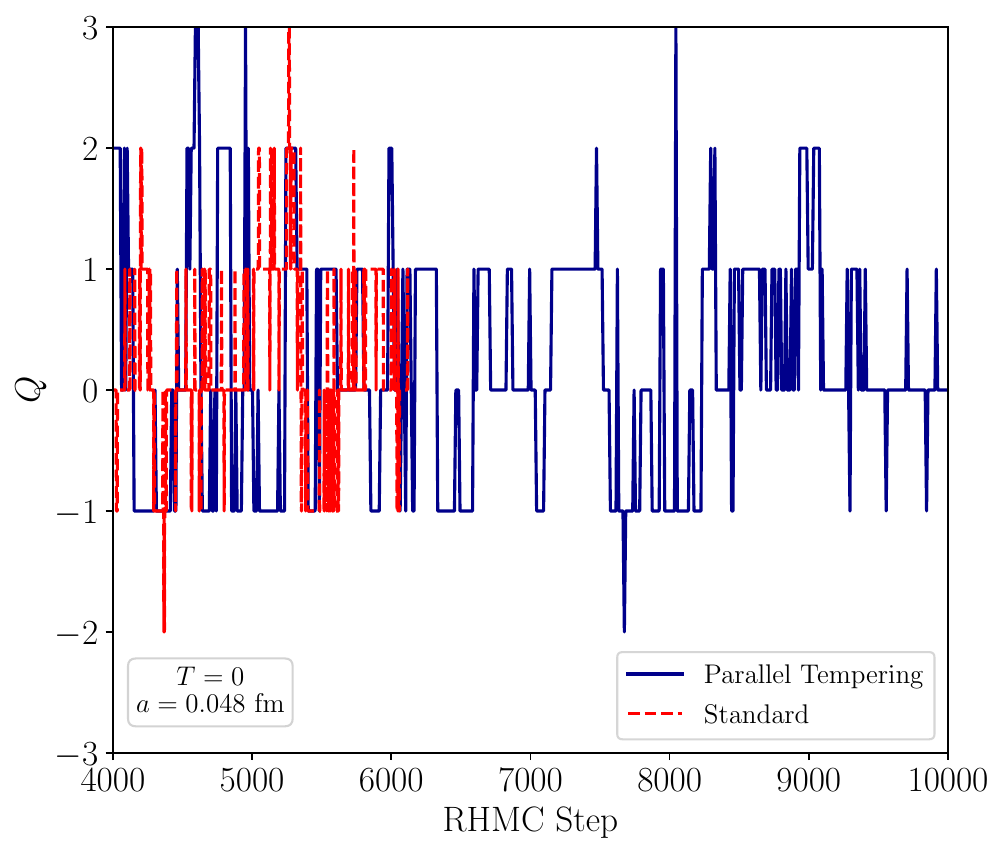}
\includegraphics[scale=0.42]{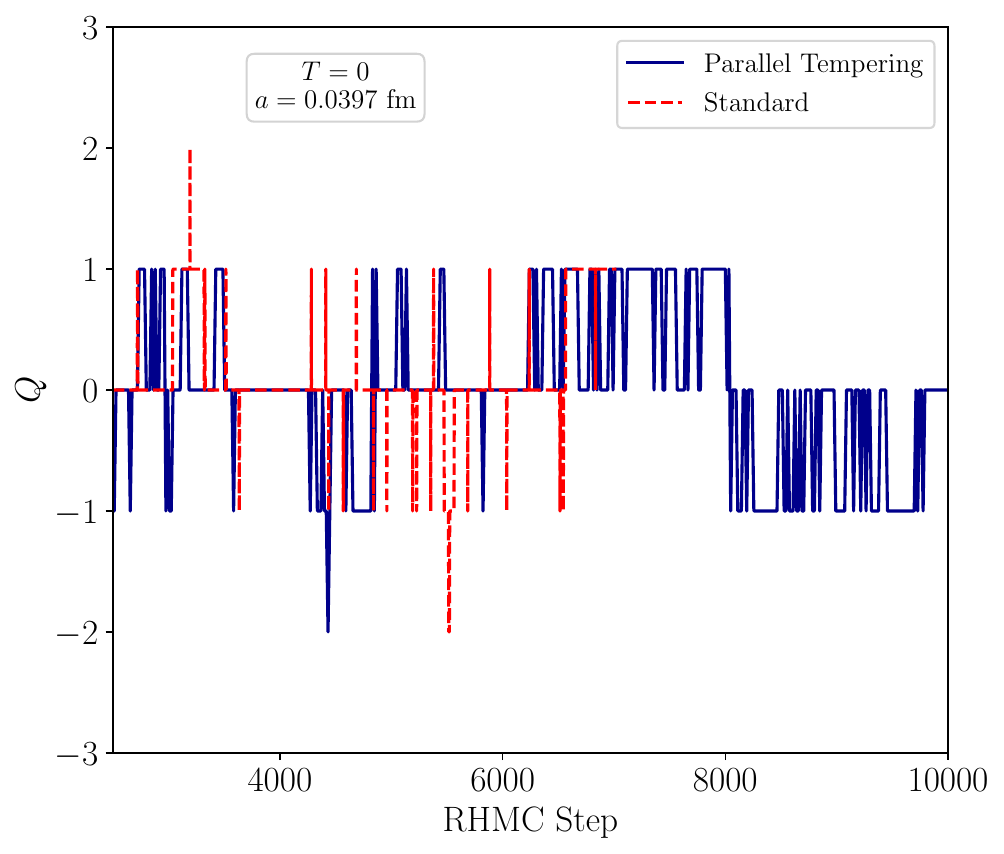}
\caption{Comparison of the Monte Carlo evolutions of the cooled lattice topological charge $Q$ obtained with the PTBC and the standard algorithm for the two lattice spacings explored at zero temperature. In both cases, the horizontal axis was reported in a common scale and expressed in numbers of RHMC steps, meaning that for PTBC we multiplied the Monte Carlo time by the number of replicas $N_r$. The parallel tempering run for the simulation point with $a=0.0480$ fm refers to the setup with $L_d=2$ and $P\approx 20\%$.}
\label{fig:historyQ_zero_T}
\end{figure}

We are now going to compare the performances of the PTBC algorithm with those of the standard algorithm. In Fig.~\ref{fig:historyQ_zero_T}, we show the Monte Carlo histories obtained with the two algorithms for the explored lattice spacings, where both Monte Carlo evolutions were expressed in units of a common $x$-scale, i.e., the number of RHMC steps, to allow for a fair comparison. This means that the Monte Carlo time of PTBC is multiplied by the number of replicas $N_r$. By a quick inspection of the histories and of the observed number of topological fluctuations, one can infer that the two algorithms seem more or less equivalent for $a=0.048$ fm, while for the finest lattice spacing $a=0.0397$ fm the improvement obtained with parallel tempering is manifest.

In order to put this comparison in more quantitative terms, it is useful to compute the integrated auto-correlation time of $Q^2$. In this study, we will use the following estimator, which allows to evaluate the integated auto-correlation time through a standard binned jack-knife analysis~\cite{Berg:2004fd,DelDebbio:2004xh}:
\beq\label{eq:tauint_def}
\tauint(Q^2) = \frac{1}{2}\left[\frac{\Delta_{\rm B}\left(Q^2\right)}{\Delta_{\rm 1}\left(Q^2\right)}\right]^2 - \frac{1}{2},
\eeq
where $\Delta_{\rm B}(Q^2)$ and $\Delta_{1}(Q^2)$ represent, respectively, the binned and the naive (i.e., obtained neglecting auto-correlations) statistical errors on $\braket{Q^2}$.\footnote{Note that the definition of the estimator of the integrated auto-correlation time given in Eq.~\eqref{eq:tauint_def} leads to $\tauint\ge 0$, with 0 the minimum value. Some authors adopt the equivalent definition without the $-1/2$, leading to $\tauint \ge 0.5$.} At the practical level, the naive error $\Delta_1$ is obtained with a unitary bin size. The binned error $\Delta_{\rm B}$ instead exhibits a plateau after a certain bin size threshold $\mathrm{B} \gtrsim \overline{\mathrm{B}}$. As discussed in Ref.~\cite{DelDebbio:2004xh}, systematics affecting this estimator are negligible when $\tauint / \mathrm{B} \ll 1$, and when $\tauint$ and $\mathrm{B}$ are both much smaller than the sample size, which we verified was always the case in our PTBC simulations. Thus, the error we assigned to the integrated auto-correlation time was assessed from the small variations observed in the plateau exhibited by $\tauint(Q^2)$ as a function of $\rm B$.

The obtained auto-correlation times are reported in Tab.~\ref{tab:res_zero_T}, again expressed in both cases in units of the number of RHMC steps, meaning that for PTBC we multiplied the value of $\tauint$ obtained from Eq.~\eqref{eq:tauint_def} by $N_r$. As previously discussed, for the $a=0.048$ fm point we find integrated auto-correlation times in agreement with each other. For the lattice spacing with $a=0.0397$ fm, instead, we find a net gain of about a factor of 2 in terms of $\tauint(Q^2)$ once the overhead introduced by the replicas is kept into account.

As a final note, we mention that our main results were obtained employing a defect size $L_d=2$, corresponding to a physical size $l_d = aL_d \sim 0.08-0.1$ fm, and tuning the tempering parameters to achieve an approximately constant $\sim 30\%$ average swap acceptance rate. However, for the simulation point with $a=0.0480$ fm, we also tried different setups. In particular, we observe that a smaller average acceptance rate $P\sim 20 \%$ for the same defect size $l_d \sim 0.1$ fm, gives perfectly compatible performances and outcomes for $\chi$, pointing out that the PTBC algorithm free parameters do not need a particular fine-tuning.

For this simulation point, we also tried a larger defect size $l_d\sim 0.2$ fm with $P\simeq 20 \%$. Also in this case we observe that the obtained result for the susceptibility is in excellent agreement with those obtained with the other parallel tempering setups. However, the integrated auto-correlation time of $Q^2$, once the overhead introduced by the replicas is taken into account, turns out to be larger compared to the standard algorithm. This is not surprising, actually it is expected. As a matter of fact, it is clear that increasing the defect size will improve the decorrelation of the topological charge in the OBC replica, but at the same time will require a large number of replicas to achieve the same fixed average swap acceptance rate (empirically, $N_r\sim L_d^{3/2}$, see also~\cite{Bonanno:2020hht,Bonanno:2024nba}), thus excessively long defects are expected to lead to worse performances compared to the standard algorithm. An analogous behavior was observed for PTBC simulations in the pure-gauge theory~\cite{Bonanno:2020hht}.

\begin{figure}[!t]
\centering
\includegraphics[scale=0.49]{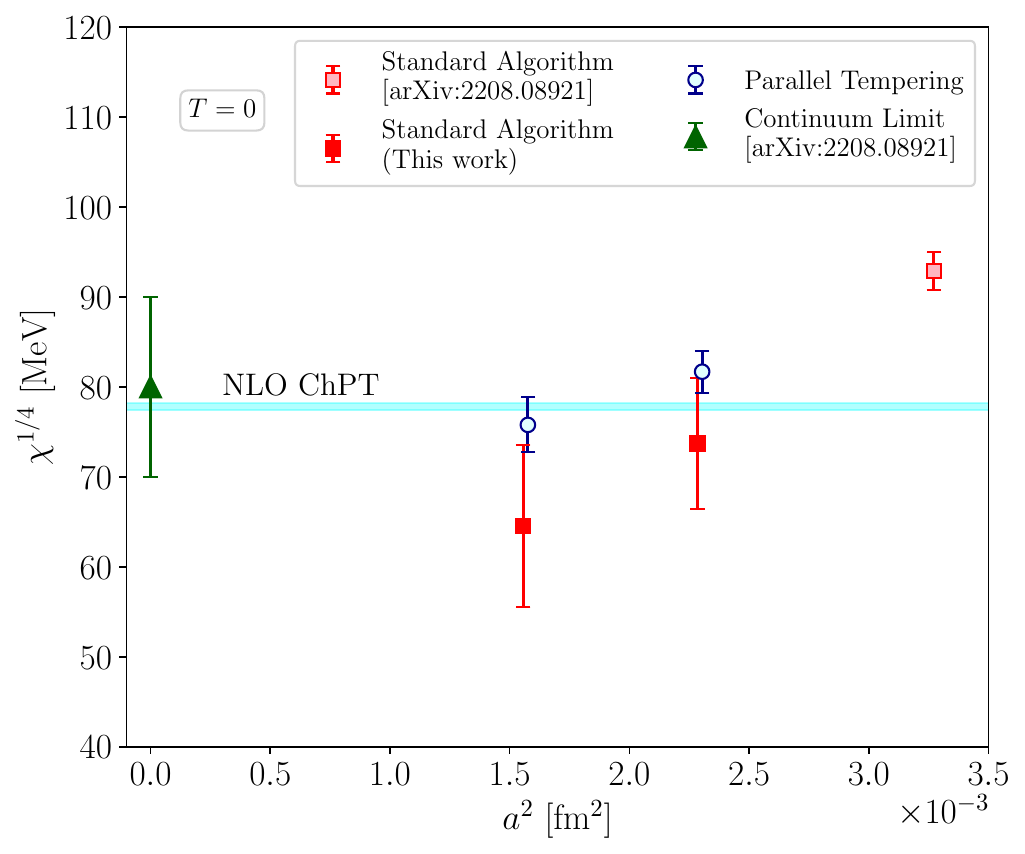}
\caption{Comparison of the lattice determinations of $\chi^{1/4}(T=0)$ obtained with the standard and the PTBC algorithms for $a=0.048$ fm and $a=0.0397$ fm with the ChPT prediction and with the continuum extrapolation obtained in Ref.~\cite{Athenodorou:2022aay} with a fermionic definition of $Q$ based on the index theorem. For the sake of comparison, we also show the result obtained for $\chi^{1/4}$ in~\cite{Athenodorou:2022aay} with the same gluonic definition considered here and for a lattice spacing $a=0.057$ fm.}
\label{fig:chi_cont_zero_T}
\end{figure}

\begin{table}[!t]
\begin{center}
\begin{tabular}{|c|c|c|c|c|c|c|}
\hline
$a$~[fm] & $P$ $(\%)$ & $L_d$ & \makecell{$\chi^{1/4}$ [MeV]\\(PTBC)} & \makecell{$\tauint(Q^2)$\\(PTBC)} & \makecell{$\chi^{1/4}$ [MeV]\\(Standard)} & \makecell{$\tauint(Q^2)$\\(Standard)}\\
\hline
\multirow{3}{*}{0.0480} & 20 & \multirow{2}{*}{2} & 81.7(2.3) & 45(15) & \multirow{3}{*}{73.7(7.3)} & \multirow{3}{*}{50(10)} \\
 & 30 & & 81.4(2.9) & 52(13) & &  \\
 & 20 & 4 & 82.8(2.8) & 108(36) & &  \\
\hline
0.0397 & 30 & 2 & 75.8(3.1) & 78(26) & 64.5(9.0) & 140(20) \\
\hline
\end{tabular}
\end{center}
\caption{Summary of the results obtained for $T\simeq 0$. The integrated auto-correlation time $\tauint(Q^2)$ was computed via Eq.~\eqref{eq:tauint_def} and is expressed for both algorithms in units of RHMC steps, meaning that the one obtained with the PTBC algorithm was multiplied by the number of replicas $N_r$.}
\label{tab:res_zero_T}
\end{table}

Finally, we show our results for the topological susceptibility, reported in Tab.~\ref{tab:res_zero_T}, in Fig.~\ref{fig:chi_cont_zero_T}, where we also compare them with the ChPT prediction in Eq.~\eqref{eq:chi_T0_ChPT}, as well as with the continuum extrapolation of~\cite{Athenodorou:2022aay}. We observe that for $a=0.048$~fm the two algorithms give consistent results for the susceptibility. The PTBC determination is more accurate but, as previously explained, once the numerical overhead introduced by the replicas is taken into account, the performances of the PTBC are essentially equivalent to those of the standard algorithm. For the finest lattice spacing explored, instead, topological freezing is significant and prevents us from obtaining a very accurate estimation of $\chi$. On the other hand, the result obtained with PTBC turns out to be perfectly consistent both with the ChPT prediction, as well as with the continuum extrapolation of~\cite{Athenodorou:2022aay}.

\subsection{Finite-temperature results}

We now move to discuss results obtained at finite temperature with the multicanonical PTBC algorithmic setup. As already explained in Sec.~\ref{sec:multican}, all finite-temperature simulations, both with and without parallel tempering, have been performed using the multicanonic method.

The finite-temperature case becomes progressively much harder than the $T=0$ one as $T$ grows because the topological susceptibility is rapidly suppressed above the chiral crossover temperature, thus the impact of the large lattice artifacts affecting $\chi$ becomes more and more important, making it harder and harder to control systematics related to the continuum extrapolation as $T$ increases, even when considering very fine lattice spacings. In Ref.~\cite{Athenodorou:2022aay}, for example, the largest temperature explored was \mbox{$T=570$ MeV $\simeq 3.7 \, T_c$.} Even though in that case the very fine range of lattice spacings $a \sim 0.057-0.029$ fm was explored, only the upper bound $\chi^{1/4}=6(6)$ MeV could be set for the topological susceptibility at that temperature using the standard gluonic discretization. Sizeable systematics were also found using a different fermionic definition of the topological charge based on spectral projectors onto the lowest-lying modes of the staggered Dirac operator, and the final continuum limit $\chi^{1/4}=8(6)$ MeV was quoted using this method, which only slightly improves on the gluonic result.

In this section we aim at re-examining this case by performing new simulations with the PTBC algorithm for the two finest lattice spacing explored in~\cite{Athenodorou:2022aay}, namely $a=0.0343$ fm and $a = 0.0286$ fm, and by also performing a new simulation for an even finer lattice spacing, namely, $a=0.0215$ fm. For these finite-$T$ simulations, we considered $N_s^3 \times N_t$ lattices with aspect ratio $N_s/N_t=4$, and again chose the bare lattice parameters so as to stay on the same LCP followed for zero-temperature runs. The summary of the simulation parameters is reported in Tab.~\ref{tab:simul_pars_T570}, along with the details of the PTBC setup.
We stress that in the thermal case, given that the lattice extensions are not isotropic, there are 
two nonequivalent ways of placing the defect, namely orthogonal or parallel to the Euclidean time direction:
in principle, one of the two could be more efficient than the other, 
however, after some preliminary test runs, we did not 
notice significant differences, and adopted a setup where the defect is orthogonal to time (i.e., the same setup of the zero-temperature case).
For the finest lattice spacing explored in the present study, we had to slightly extrapolate the bare parameters of the LCP of Refs.~\cite{Aoki:2009sc, Borsanyi:2010cj, Borsanyi:2013bia} via an educated guess rooted in perturbation theory predictions, and more details on this can be found in App.~\ref{app:LCP_extr} (note, though, that this is completely irrelevant for the sake of comparing the algorithmic performances of the PTBC with respect to the standard multicanonic algorithm).

\begin{table}[!t]
\begin{center}
\begin{tabular}{|c|c|c|c|c|c|c|c|}
\hline
$\beta$ & $a$~[fm] & $am_s \cdot 10^{2}$ & $N_s$ & $N_t$ & $N_r$ & $L_d$ & $P$ $(\%)$\\
\hline
\multirow{2}{*}{4.459} & \multirow{2}{*}{0.0343} & \multirow{2}{*}{1.370} & \multirow{2}{*}{40} & \multirow{2}{*}{10} & 7 & 1 & 40(10)\\
&&&&& 10 & 2 & 19(4)\\
\hline
4.592 & 0.0286 & 1.090 & 48 & 12 & 10 & 2 & 19(4)\\
\hline
4.798 & 0.0215 & 0.798 & 64 & 16 & 10 & 2 & 20(4)\\
\hline
\end{tabular}
\end{center}
\caption{Summary of simulation parameters for the $T\simeq 570$ MeV runs, performed on  $N_s^3 \times N_t$ lattices. The bare parameters $\beta$ and $am_s$ and
the lattice spacings $a$ have been fixed according to the LCP determined in Refs.~\cite{Aoki:2009sc, Borsanyi:2010cj, Borsanyi:2013bia} with $m_\pi = 135$ MeV, and
$am_l$ is fixed through the physical strange-to-light quark mass ratio $m_s/m_l=28.15$ (see the text and App.~\ref{app:LCP_extr} for more details). The value of $P$ represents the average value of the swap probabilities, and the dispersion around it.}
\label{tab:simul_pars_T570}
\end{table}

\begin{figure}[!t]
\centering
\includegraphics[scale=0.4]{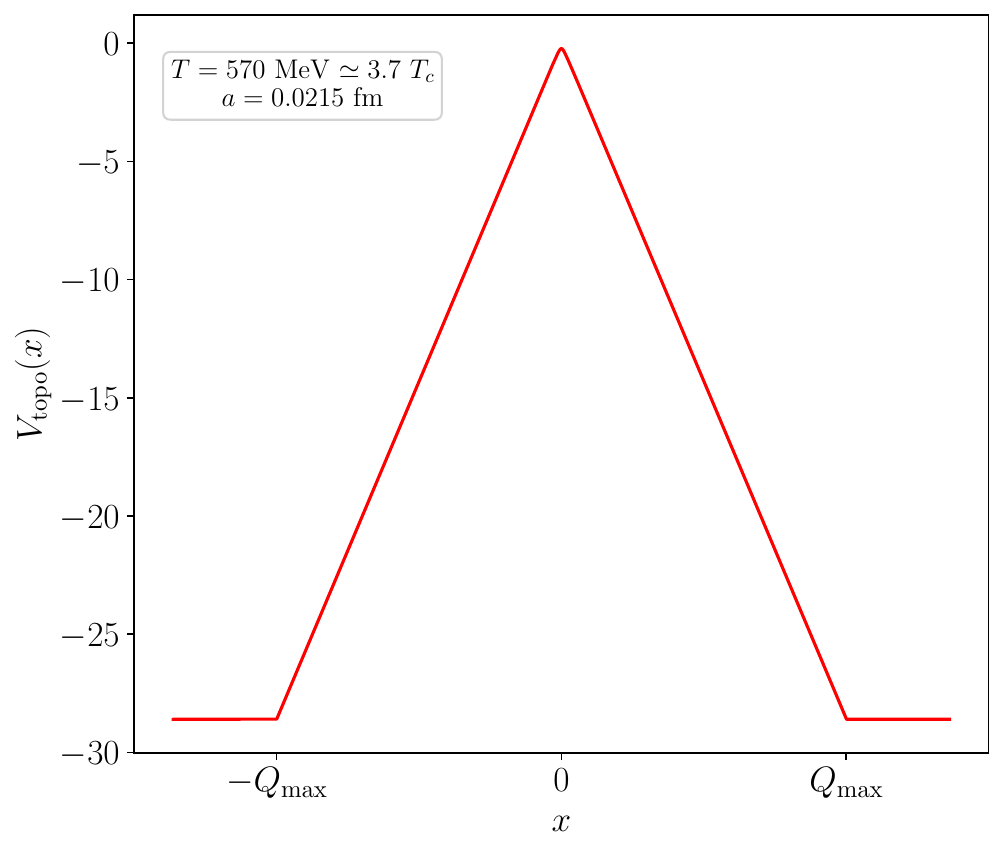}
\caption{Example of the shape of the topological bias potential used in our multicanonical simulations with $a=0.0215$ fm. The adopted functional forms are the same already employed in Refs.~\cite{Bonati:2018blm,Athenodorou:2022aay}: $V_{\rm topo}(x) = -\sqrt{(Bx)^2+C)}$ for $\vert x \vert < Q_{\rm max}$ and $V_{\rm topo}(x) = -\sqrt{(BQ_{\rm max})^2+C)}$ for $\vert x \vert > Q_{\rm max}$. In this case $B=13$, $C=0.05$ and $Q_{\rm max}=2$.}
\label{fig:topot_ex}
\end{figure}

\begin{table}[!t]
\begin{center}
\begin{tabular}{|c|c|c|c|c|c|}
\hline
$a$~[fm] & $L_d$ & \makecell{$\chi^{1/4}$ [MeV]\\(PTBC)} & \makecell{$\tauint(Q^2)$\\(PTBC)} & \makecell{$\chi^{1/4}$ [MeV]\\(Standard)} & \makecell{$\tauint(Q^2)$\\(Standard)}\\
\hline
\multirow{2}{*}{0.0343} & 1 & 28.2(1.3) & 10(3) & \multirow{2}{*}{29.9(2.3)} & \multirow{2}{*}{30(12)} \\
 & 2 & 27.4(1.1) & 65(15) &  &  \\
\hline
0.0286 & 2 & 19.22(71) & 27(4) & 19.5(3.2) & 50(10) \\
\hline
0.0215 & 2 & 14.44(77) & 26(6) & 14(3)$^*$ & $\gtrsim 10^2$ \\
\hline
\end{tabular}
\end{center}
\caption{Summary of the results obtained for $T\simeq 570$ MeV. The results obtained for $a=0.0343$ fm and $0.0286$ fm with the standard multicanonic algorithm are taken from Ref.~\cite{Athenodorou:2022aay}. The integrated auto-correlation time $\tauint(Q^2)$ was computed via Eq.~\eqref{eq:tauint_def} and is expressed for both algorithms in units of RHMC steps, meaning that the one obtained with the PTBC algorithm was multiplied by the number of replicas $N_r$. The $^*$ points out that, in that run with the standard algorithm, our estimate of $\tauint(Q^2)$ is very rough, thus, the statistical error on $\chi$ is very likely to be underestimated.}
\label{tab:res_T570}
\end{table}

\begin{figure}[!t]
\centering
\includegraphics[scale=0.4]{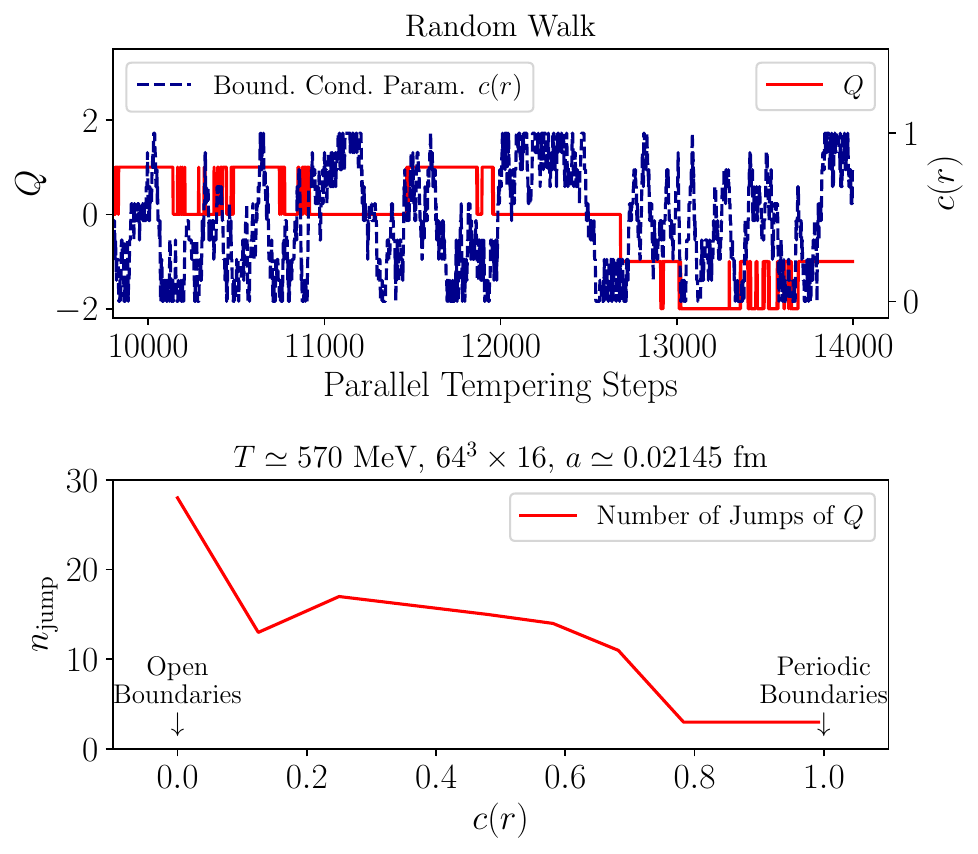}
\includegraphics[scale=0.4]{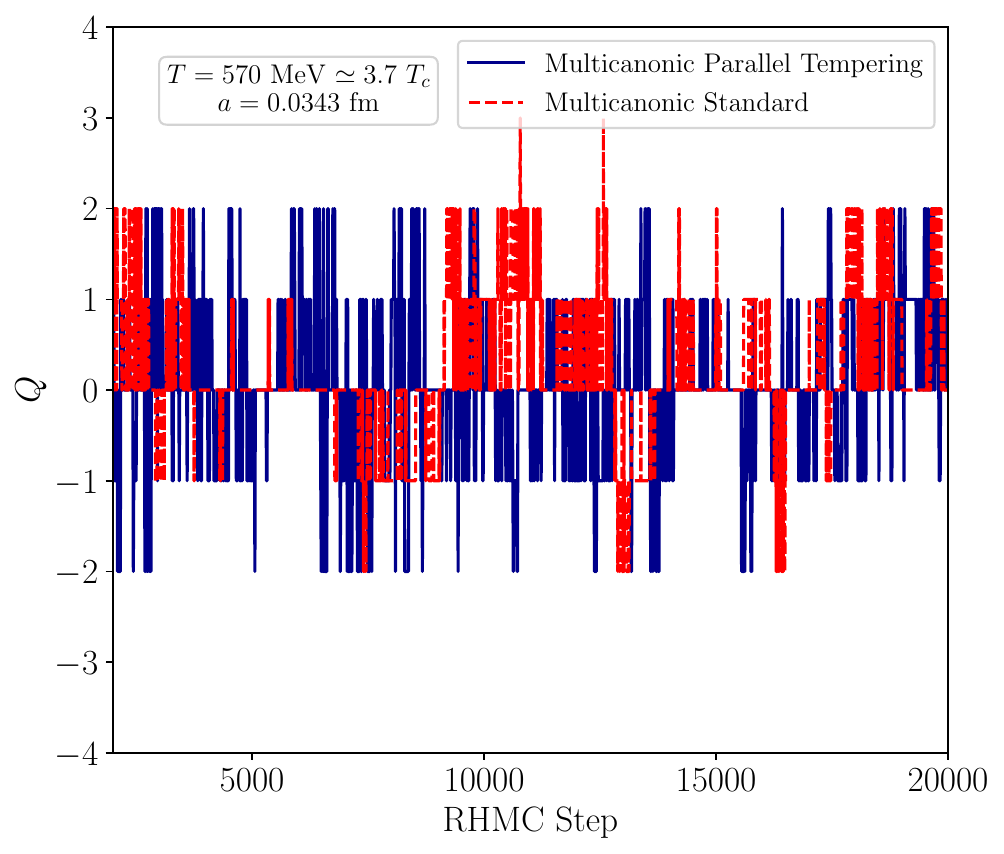}
\includegraphics[scale=0.4]{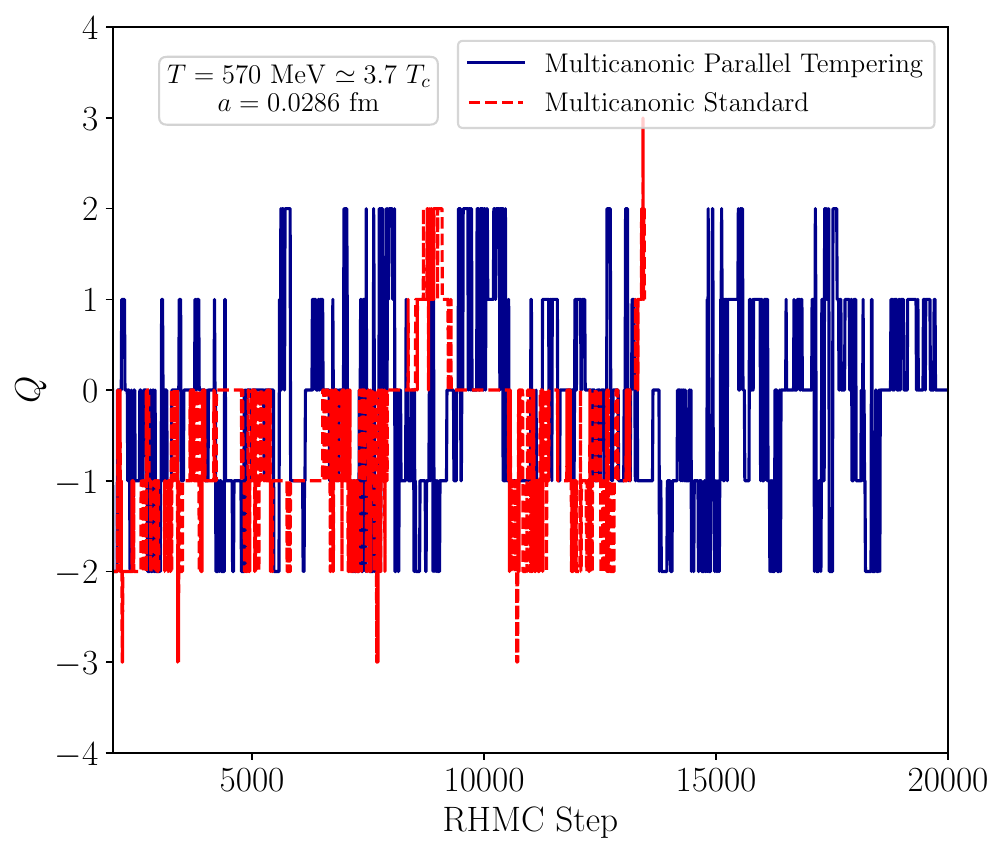}
\includegraphics[scale=0.4]{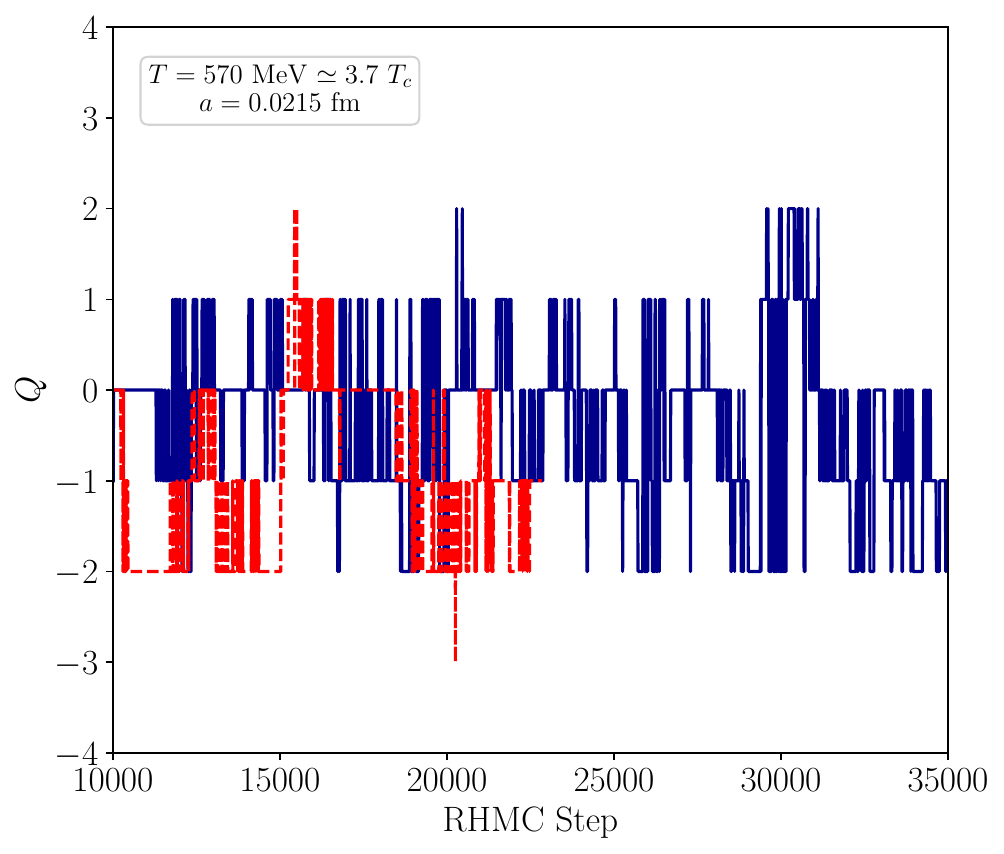}
\caption{Top left plot: random walk of a configuration among the different replicas. We display how the boundary condition parameter $c(r)$ and the topological charge $Q$ change as a function of the Monte Carlo time, as well as the number of jumps of the topological charge as a function of $c(r)$. Top right and bottom plots: comparison of the multicanonical Monte Carlo evolutions of the cooled lattice topological charge $Q$ obtained with the PTBC and the standard algorithm for all lattice spacings explored at $T=570$ MeV. In all cases, the horizontal axis was reported in a common scale and expressed in numbers of RHMC steps, meaning that for PTBC we multiplied the Monte Carlo time by the number of replicas $N_r$. The parallel tempering run for the simulation point with $a=0.0343$ fm refers to the setup with $L_d=1$.}
\label{fig:comp_storyQ_T570}
\end{figure}

We start our discussion by comparing the efficiency of the PTBC and the standard multicanonic algorithms. In Fig.~\ref{fig:topot_ex}, we plot an example of the multicanonic bias potential employed to enhance the probability of visiting suppressed topological sectors, while in Fig.~\ref{fig:comp_storyQ_T570} we show the Monte Carlo evolutions of the lattice topological charge $Q$ obtained with both algorithms, again compared by reporting both histories on a common horizontal axis (which implies that the Monte Carlo time of PTBC has been multiplied by the number of replicas $N_r$). From a visual inspection of the Monte Carlo histories, we observe that approaching the continuum limit parallel tempering outperforms the standard algorithm.

This is in particular clearly exemplified by the top left plot of Fig.~\ref{fig:comp_storyQ_T570}, where we show how the topological charge $Q$ changes while a given configuration performs its random walk through the different replicas, experiencing different boundary conditions, parametrized by $c(r)$. As it can be seen, the topological charge is more likely to jump between two different sectors when $c(r)$ is closer to 0 (OBCs), and less likely when $c(r)$ is closer to 1 (PBCs). This means that the relevant source of topological fluctuations observed in the PTBC Monte Carlo histories is due to the opening of the boundary conditions, and not just a mere illusion introduced by the swaps among different replicas. To put this observation on a more quantitative ground, we computed the Pearson correlation coefficient $C=\left(\braket{xy}-\braket{x}\braket{y}\right)/\left(\sqrt{\braket{x^2}-\braket{x}^2} \sqrt{\braket{y^2}-\braket{y}^2}\right)$ between the boundary condition coefficient $x_i = c_i(r)$ and the jump of the topological charge $y_i=\vert \Delta Q_i \vert = \vert Q_{i+1} - Q_i\vert$, with $i$ the PTBC step, finding $C=-0.072(18)$. This result quantifies that it exists a non-vanishing correlation between the jumps of the topological charge and the opening of the boundaries on the defect region.

In order to better quantify the improvement obtained with PTBC with respect to the standard algorithm, we computed the integrated auto-correlation time of $Q^2$, expressed in both cases using the same units (i.e., the number of RHMC steps). Results are reported in Tab.~\ref{tab:res_T570}. It is important to stress here that these integrated auto-correlation times refer to Eq.\eqref{eq:tauint_def} calculated using the errors on $\braket{Q^2}$ obtained \emph{after} reweighting. The motivation for this choice stems from the fact that we want a definition of $\tauint(Q^2)$ which directly reflects the statistical accuracy on our target quantity (namely, the susceptibility after reweighting the topological bias potential away).\footnote{For a related discussion on $\tauint(Q^2)$ in the context of Jarzinsky-inspired reweighting in out-of-equilibrium simulations see Ref.~\cite{Bonanno:2024udh}.}

For the coarsest lattice spacing explored with the PTBC algorithm, $a=0.0343$ fm, we tried two different defect lengths, namely $L_d=1$ and $2$. In the latter case, the PTBC auto-correlation time (i.e., multiplied by $N_r$) turned out to be larger by about a factor of 2 with respect to the standard algorithm one. In the former case instead, we observe substantial better performances compared to the standard algorithm, obtaining a net gain of about a factor of 3 in terms of $\tauint(Q^2)$. This thus suggests that defects smaller than $a L_d \sim 0.07$ fm should be employed in this case. Thus, for the other runs at finer lattice spacings, we simply adopted $L_d=2$ in all cases.

For what concerns $a=0.0286$ fm, we observed a net gain of a factor of $2$ in terms of computational power. This has allowed us to greatly reduce the error on the gluonic determination for this lattice spacing compared to the result of Ref.~\cite{Athenodorou:2022aay}. For the finest lattice spacing explored the gain attained with PTBC algorithm is even more impressive. Indeed, while for parallel tempering we find an auto-correlation time which is of the same order of that observed for the coarser simulation points explored, using the standard multicanonic algorithm we are only able to make a rough estimate of $\tauint(Q^2)$. Interestingly, we find that, with the standard algorithm, the integrated auto-correlation time turns out to be larger by at least a factor of 4 compared to PTBC. Thus, these results point out that the PTBC algorithm shows an improved scaling of the auto-correlation time as a function of $1/a$ with respect to the standard algorithm also in the finite-temperature case, analogously to what has been deduced from our zero-temperature runs.

As a final comment, we observe that for our high-temperature runs defect sizes corresponding to a physical length $ l_d = a L_d \sim 0.034-0.057$ fm turned out to be enough to obtain much better performances compared to the standard algorithm. These defect sizes are about a factor of 2 smaller than those employed at zero temperature. A possible explanation of this observation can be the following. As it is well-known from DIGA standard arguments, at high temperature the contribution of large instantons with size $\rho\gg 1/T$ to the path integral is suppressed. Since one can expect the topological excitations created/annihilited through the defect to have a typical size $\rho \sim l_d$, one can expect to need smaller defects (in physical units) at high temperatures to obtain satisfying performances with the PTBC algorithm.\footnote{Such argument could also explain why, at finite temperature, topological freezing seems to kick in at finer lattice spacings compared to the zero-temperature case. Indeed, the path integral at high $T$ is dominated by the contribution of small instantons, which are more easily created/annihilated adopting local updating algorithms. A similar behavior was observed, e.g., in $2d$ $\mathrm{CP}^{N-1}$ models for small values of $N$, whose topological dynamics, unlike in the large-$N$ case, is dominated by small instantons, and is found to exhibit smaller auto-correlation times for the topological charge~\cite{Berni:2020ebn,Bonanno:2022dru}.}

\begin{table}[!t]
\begin{center}
\begin{tabular}{|c|c|}
\hline
\multicolumn{2}{|c|}{$T=570$ MeV}\\
\hline
$a$~[fm] & $\chi^{1/4}$ [MeV] \\
\hline
0.0429 & 36.7(2.6) \\
0.0343 & 27.4(1.1) \\
0.0286 & 19.22(71) \\
0.0215 & 14.44(77) \\
0      & 6.2(1.2)  \\
\hline
\end{tabular}
\end{center}
\caption{Determinations of $\chi^{1/4}$ MeV at finite lattice spacing employed to perform the continuum extrapolation assuming leading $O(a^2)$ corrections. The coarsest lattice spacing determination comes from the gluonic results of Ref.~\cite{Athenodorou:2022aay}. The determination for $a=0$ is the final continuum extrapolation obtained in this work.}
\label{tab:contlimit_T570}
\end{table}

\begin{figure}[!t]
\centering
\includegraphics[scale=0.45]{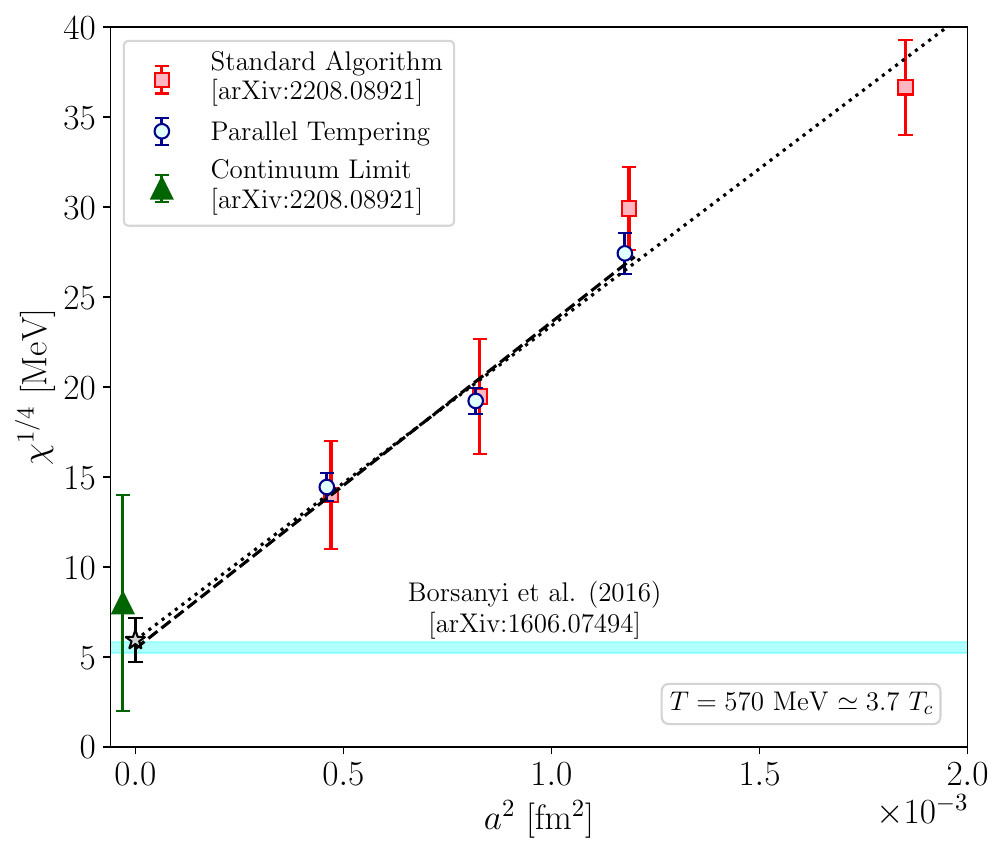}
\caption{Continuum extrapolation of the determinations of $\chi^{1/4}$ obtained for $T=570$ MeV reported in Tab.~\ref{tab:contlimit_T570}, compared to the continuum limits obtained in Refs.~\cite{Borsanyi:2016ksw,Athenodorou:2022aay}.}
\label{fig:contlimit_T570}
\end{figure}

We can now add the new data obtained with the PTBC algorithm to the results obtained at coarser lattice spacings in~\cite{Athenodorou:2022aay} in order to perform a continuum extrapolation of these determinations. The data employed for the continuum limit can be found in Tab.~\ref{tab:contlimit_T570}, while the continuum extrapolation is displayed in Fig.~\ref{fig:contlimit_T570}. Assuming standard leading $O(a^2)$ correction to the continuum limit, we observe that a linear fit in $a^2$ including all 4 available points yields the continuum value:
\beq\label{eq:final_res_T570}
\chi^{1/4} = 6.2(1.2)\text{ MeV}, \qquad (T=570\text{ MeV}).
\eeq
Excluding the coarsest point yields the perfectly compatible result $6.0(1.5)$ MeV. Thus, thanks to the addition of the new PTBC data, we are able to perfectly control systematic errors coming from the continuum limit extrapolation, and we can simply take Eq.~\eqref{eq:final_res_T570} as our final result for $\chi^{1/4}(T=570\text{ MeV})$. We also observe that our result greatly improves on the result of Ref.~\cite{Athenodorou:2022aay}, $8(6)$ MeV, and is also perfectly in agreement with the result obtained in Ref.~\cite{Borsanyi:2016ksw} for the same temperature: $5.5(3)$ MeV, cf.~also Fig.~\ref{fig:contlimit_T570}.
To be fair, a systematic error should be added to our determination, in relation to the scale setting for the finest lattice spacing, which has been fixed only by extrapolation from other scale settings; however, this is not relevant in order to assess to improvement achieved by the new algorithm, which is expressed solely by the improved accuracy of the result in Eq.~\eqref{eq:final_res_T570}.

\section{Conclusions}\label{sec:conclu}

In this manuscript, we have presented a first investigation of the efficiency of a novel numerical algorithm to deal with the computational problem of topological freezing in full QCD simulations involving dynamical fermions, namely, the Parallel Tempering on Boundary Conditions algorithm. This method, initially proposed in $2d$ $\mathrm{CP}^{N-1}$ by M.~Hasenbusch and soon after implemented and successfully employed in $4d$ $\mathrm{SU}(N)$ pure-gauge theories, is here applied to $N_f=2+1$ full QCD simulations at the physical point, both at zero and finite temperature, for the first time.

The main idea underlying this algorithm is to combine periodic boundary conditions and open boundary conditions in a parallel tempering framework in order to exploit the improved scaling of the auto-correlation time as a function of the inverse lattice spacing achieved with open boundaries, while at the same time avoiding unphysical boundary effects, as all observables are computed on the periodic replica, where translation-invariance, and thus a proper notion of global topological charge, is kept.

We have performed simulations of $2+1$ QCD with physical quark masses for very fine lattice spacings $a \sim 0.048 - 0.021$ fm, both at zero and finite temperature, adopting rooted stout-staggered fermions to discretize the quark action, although this is not a mandatory choice, as our present implementation of the PTBC algorithm can be easily adapted to other fermion discretizations. We found that the PTBC algorithm yields an auto-correlation time for $Q^2$ which is generally much smaller than the one attained with standard RHMC simulations, allowing in all cases to improve our previous estimates of the topological susceptibility. Moreover, in both cases we have found perfect agreement in the obtained results for $\chi$ after comparing with previous results reported in the literature and/or with analytic predictions when available. Finally, we observed that no hard fine-tuning of the parameters of the PTBC algorithm is required to obtain optimal performances. In practice, we observed that defect sizes of the order of $\sim 0.08-0.1$ fm at zero temperature and $\sim 0.03 - 0.06$ fm at finite temperature, with $\sim O(10)$ replicas and an avarage $\sim 20 - 30 \%$ swap acceptance rates, were largely sufficient to obtain a perfectly working setup with exceedingly better performances compared to the standard algorithm. Concerning the results obtained for the topological susceptibility, we have been able to greatly improve on the determinations obtained in Ref.~\cite{Athenodorou:2022aay} both at $T=0$ and at $T=570$ MeV $\simeq 3.7 \, T_c$.

Given the enhancements obtained from this first investigation of the efficiency of the method, the PTBC algorithm can open up several new research paths that we would like to explore in the near future. For example, it would be extremely interesting to use this algorithm to improve past studies of the temperature-behavior of the topological susceptibility~\cite{Athenodorou:2022aay} or of the QCD sphaleron rate~\cite{Bonanno:2023ljc,Bonanno:2023thi} in the chirally-restored high-temperature phase of QCD, and in particular to extend previous investigations in order to reach the GeV scale. Of course, this requires first to perform and extend presently available scale settings in order to reach much finer lattice spacings of the order of $a\sim 0.01$ fm, which can now be efficiently done with the PTBC algorithm using, e.g., the finite-temperature techniques put forward in Ref.~\cite{Junnarkar:2023fal}.

\section*{Acknowledgements}
The work of C.~B.~is supported by the Spanish Research Agency (Agencia
Estatal de Investigaci\'on) through the grant IFT Centro de Excelencia Severo
Ochoa CEX2020-001007-S and, partially, by grant PID2021-127526NB-I00, both
funded by MCIN/AEI/ 10.13039/ 501100011033. C.~B.~also acknowledges
support from the project H2020-MSCAITN-2018-813942 (EuroPLEx) and the EU
Horizon 2020 research and innovation program, \\STRONG-2020 project, under
grant agreement No 824093. G.~C.~acknowledges support from the National Centre on HPC, Big Data and Quantum Computing -- SPOKE 10 (Quantum Computing) and received funding from the European Union Next-GenerationEU -- National Recovery and Resilience Plan (NRRP) -- MISSION 4 COMPONENT 2, INVESTMENT N.~1.4 – CUP N.~I53C22000690001. L.~M.~acknowledges support by the French Centre national de la recherche scientifique (CNRS) under an Emergence@INP 2023 project. This work has also been partially supported by the project ``Non-perturbative aspects of fundamental interactions, in the Standard Model and beyond'' funded by MUR, Progetti di Ricerca di Rilevante Interesse Nazionale (PRIN), Bando 2022, grant 2022TJFCYB (CUP I53D23001440006).
Numerical calculations have been performed on the
\texttt{Leonardo} machine at Cineca, based on the agreement between INFN and
Cineca, under projects INF23\_npqcd and INF24\_npqcd.

\appendix

\section{Extrapolation of bare parameters to extend the physical point Line of Constant Physics}\label{app:LCP_extr}

\begin{figure}[!t]
\centering
\includegraphics[scale=0.42]{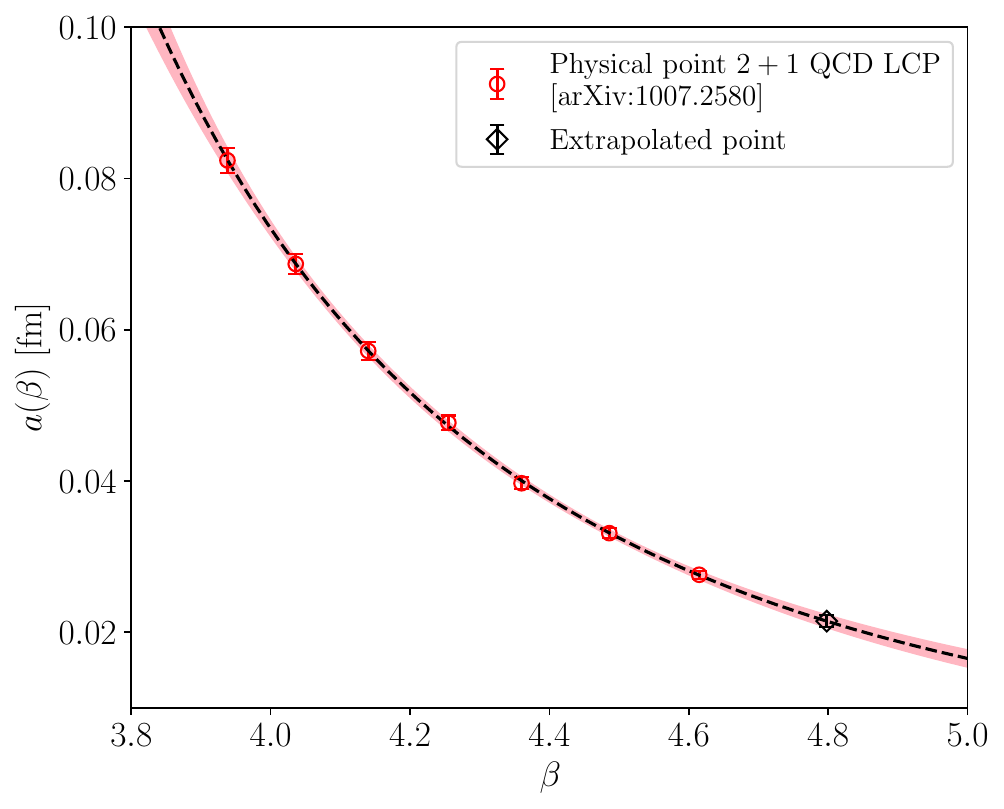}
\includegraphics[scale=0.42]{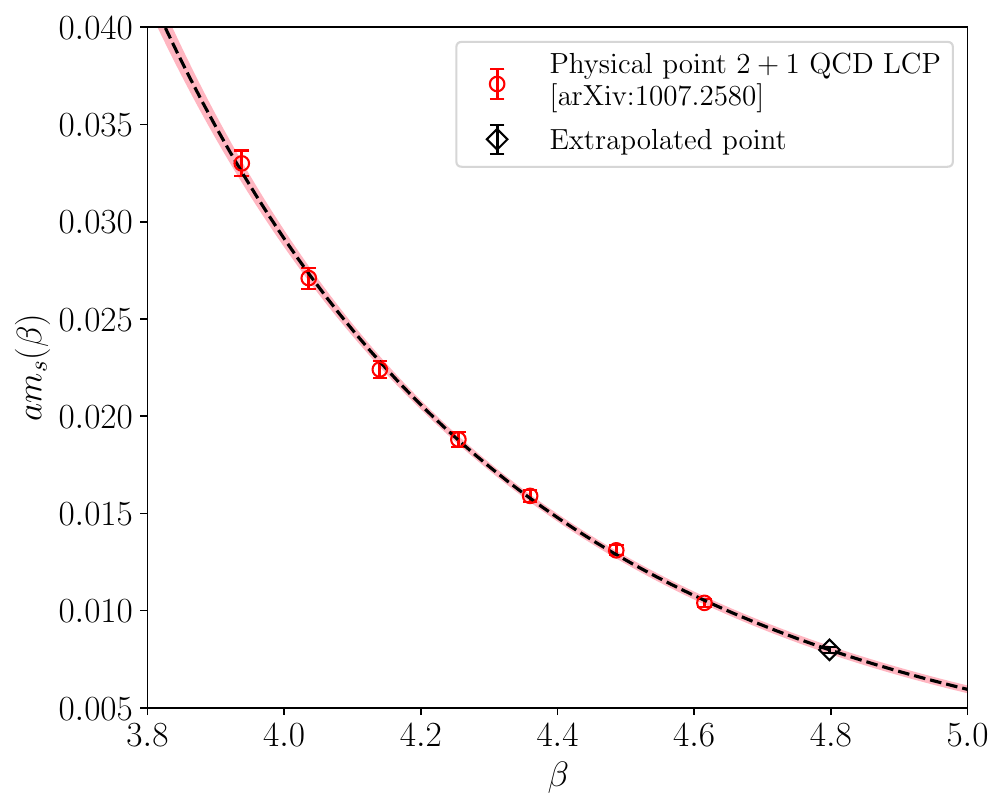}
\caption{Extrapolation of the LCP determined in Refs.~\cite{Aoki:2009sc, Borsanyi:2010cj, Borsanyi:2013bia} to determine the bare parameters corresponding to $a\simeq 0.215$ fm in Tab.~\ref{tab:simul_pars_T570}, see the text in App.~\ref{app:LCP_extr} for more details.}
\label{fig:LCP_extr}
\end{figure}

In order to extend the LCP determined in Refs.~\cite{Aoki:2009sc, Borsanyi:2010cj, Borsanyi:2013bia} to the lattice spacing $a=0.0215$ fm, we performed a best fit of $a(\beta)$ and $am_s(\beta)$ using educated guesses based on the expected perturbative behaviors of these quantities. For what concerns the lattice spacing, we used the same fit function proposed and employed in Eq.~(35) of Ref.~\cite{Lucini:2005vg}, suitably adapted from the $N_f=0$ to the $N_f=2+1$ case using the well-known universal coefficients of the QCD beta-function. For the bare strange quark mass, we instead used a fit function modeled on the basis of the expected leading perturbative behavior of the quark mass anomalous dimension. In particular, we used the following fit function: $am_s(\beta)=A\beta^{-B}$ (with $A$ and $B$ free fit parameters), see, e.g., Eq.~(2.9) of Ref.~\cite{DellaMorte:2005kg}. The resulting best fits to the known LCP points, along with the best fit curves and the corresponding extrapolated bare parameters, are shown in Fig.~\ref{fig:LCP_extr}. Finally, $am_l$ was obtained from $m_s/m_l=28.15$.

\FloatBarrier

\FloatBarrier

\providecommand{\href}[2]{#2}\begingroup\raggedright\endgroup

\end{document}